\documentclass[10pt]{article}
\usepackage{amssymb,euscript,bbold}
\usepackage{amsfonts}
\usepackage{amssymb}
\usepackage{mathrsfs}
\usepackage{graphicx}
\usepackage{slashed}
\usepackage{amsthm}
\usepackage{amsmath}
\UseRawInputEncoding
\newcommand{\be}{\begin{equation}}
\newcommand{\ee}{\end{equation}}
\newcommand{\ba}{\begin{eqnarray}}
\newcommand{\ea}{\end{eqnarray}}

\newcommand{\ft}{\footnote}

\newcommand{\op}[1]{\operatorname{#1}} %shorter command for typesetting operators
\usepackage{xcolor}
\usepackage{bbold} % To get nice looking blackboard bold 1's
\usepackage{hyperref}

\begin{document}
\input{epsf}

\begin{flushright}
KCL-PH-TH/2023-23
\end{flushright}
\begin{flushright}
%{\sf \today}
\end{flushright}
\begin{center}
\Large{Coulomb and Higgs Phases of $G_2$-manifolds.}\\
\bigskip
\large{B.S. Acharya} and 
\large{D.A. Baldwin}\\
%\ft{}\\
\smallskip\normalsize{\it
Abdus Salam International Centre for Theoretical Physics, Strada Costiera 11, 34151, Trieste, Italy}\\

and

{\it Department of Physics, Kings College London, London, WC2R 2LS, UK}\\
\end{center}

%\renewcommand{\abstractname}{\sc Abstract}
%\begin{Abstract}
\bigskip
\begin{center}
{\bf {\sc Abstract}}
Ricci flat manifolds of special holonomy are a rich framework as models of the extra dimensions in string/$M$-theory. At special points in vacuum moduli space, special kinds of singularities occur and demand a physical interpretation. In this paper we show that the topologically distinct $G_2$-holonomy manifolds arising from desingularisations of codimension four orbifold singularities due to Joyce and Karigiannis correspond physically to Coulomb and Higgs phases of four dimensional gauge theories. The results suggest generalisations of the Joyce-Karigiannis construction to arbitrary ADE-singularities and higher order twists which we explore in detail in explicitly solvable local models. These models allow us to derive an isomorphism between moduli spaces of Ricci flat metrics on these non-compact $G_2$-manifolds and flat ADE-connections on compact flat 3-manifolds which we establish explicitly for $\op{SU}(n)$.   
\end{center}
%\bigskip
%\normalsize

%\end{center}
%\end{abstract}
\newpage

\section{Introduction.}

One of the important lessons from superstring/$M$-theory over the last four decades has been the significant role played by special kinds of singularities in space. The broad and rich framework of the underlying theory can often be used to show that certain singularities may be perfectly sensible physically and, moreover, often support localised, light, interacting degrees of freedom. The geometric and topological properties of such singularities often provide microscopic insights into the fundamental properties of the quantum field theories which describe these degrees of freedom. Therefore it is important to try to understand which kinds of singularities are physically sensible and to provide a description of the physics supported at such singularities.

The classic examples of such singularities are orbifold singularities in space \cite{Stringsonorbifolds1, Stringsonorbifolds2} of which the supersymmetric, conical ADE-singularities ($\mathbb{C}^{2}/\Gamma_{ADE}$) are perhaps the best understood. Other supersymmetric cases are also reasonably well understood to some extent, such as singularities in Calabi-Yau threefolds, but there is much more to explore; for instance 
much of the literature focuses on algebraic descriptions of such singularities whilst the properties of the spacetime background are less well studied. See \cite{acharya2022-5dSCFTs} for further comments.

We will discuss four dimensional supersymmetric vacua of $M$-theory obtained by modelling the 7 extra dimensions by a space $X$, with metric $g$, whose holonomy group is the exceptional Lie group $G_2$. 
In particular, for smooth $X$, the physics of $M$-theory has only Abelian gauge symmetries and only neutral light particles. Non-Abelian gauge fields have been shown to arise from codimension four orbifold singularities of $X$ \cite{bsa2, acharya1999m} whilst chiral fermions arise from particular conical codimension seven singularities \cite{ew2}. In both cases, these degrees of freedom are in fact wrapped $M$2-branes which have collapsed to formally zero size (and mass) at the singularity. Such models of $M$-theory on $G_2$-holonomy spaces have been shown to give rise to models of physics beyond the Standard Model with a rich phenomenology \cite{G2-mssm, nonthermalDM}.

Proving the existence of $G_2$-holonomy metrics on a compact 7-manifold $X$ is notoriously difficult. The known existence results involve surgery and gluing methods whereby one constructs $X$ by gluing together non-compact model spaces along common boundary regions \cite{joyce1996compact1,joyce1996compact2,kovalev2003twisted,CHNP,Nordstrom}. Often, one starts with a model space $X_0$ that itself has very special kinds of singularities, removes a neighbourhood of the singular regions and glues in a suitable model space which gives a smooth $X$. One then uses perturbation theory methods to prove the existence of the $G_2$-holonomy metric \cite{joyce1996compact1,dominic}.  Thankfully, the kinds of singularities which arise in gluing constructions themselves tend to have interesting and sensible physical interpretations, with localised light degrees of freedom, often describable by an interacting quantum field theory.

Joyce and Karigiannis \cite{joyce2021new} have shown that, under certain topological assumptions, that perhaps the simplest orbifold singularities in a compact $G_2$-holonomy space $(X_0, g_0)$ can be desingularised to produce smooth topologically distinct $G_2$-manifolds $(X_c, g_c)$ and $(X_h, g_h)$ respectively. The goal of this paper is to interpret these results physically and to generalise them to more complicated singularities. The main conclusions are that the topologically distinct desingularisations considered by Joyce and Karigiannis are describable physically by the Coulomb and Higgs branches, respectively, of certain four dimensional gauge theories. The basic result is explained in section three, after reviewing the relevant features of Joyce-Karigiannis in section two.

In section four we introduce some simple, exactly solvable local models of the kind originally introduced in \cite{acharya1999m} and subsequently studied in \cite{Barbosa:2019bgh, ReidegeldThesis}. These models are obtained as fibrations of ADE-type ALE (or even ALF) spaces over compact flat 3-manifolds and we establish a correspondence 
between the (complexified) moduli space of $G_2$-holonomy metrics on these 7-manifolds, the moduli space of (complex) flat connections over the 3-manifolds and the classical moduli space of the physical four dimensional field theories.
At the end of the paper we combine all the results to show that massless matter in fundamental representations of the gauge
group arise at special points in the semi-classical moduli space of $M$-theory on certain {\it compact} $G_2$-holonomy manifolds and determine the light particle spectrum of most of Joyce's compact examples.

{\it Background Material and Notation.}
In this paragraph, for the ease of the reader, we collect some background material and definitions concerning $G_2$-manifolds.
A $G_2$-structure on a compact 7-manifold $X$, is defined by a 3-form, $\varphi$ which is $G_2$-invariant at each point wrt the natural action of $G_2$ on the tangent spaces of $X$ at each point: $\mathbb{R}^{7} \equiv T(X)|_{pt}$. Since $G_2 \subset \op{SO}(7)$, $\varphi$ induces a metric, $g(X)$ and orientation on $X$. The holonomy group of $g(X)$ is a subgroup of $G_2$ if and only if $\varphi$ is parallel wrt the Levi-Cevita connection, $\nabla_g \varphi=0$ . This is equivalent to $d\varphi=d^*\varphi = 0$. If $\nabla_g \varphi=0$ and the universal cover of $X$ is compact then $Hol(g(X))=G_2$. The latter conditions are equivalent to the existence of a single parallel spinor field, $\eta$, $\nabla_g \eta = 0$, which also implies that $G_2$-manifolds preserve supersymmetry when used as models of the extra dimensions in superstring/$M$-theory. Since the Ricci tensor of a $G_2$-holonomy metric is identically zero, the classical vacuum of such models have zero cosmological constant.

\section{Joyce-Karigiannis Manifolds.}

In this section we give a brief overview of constructions of compact $G_2$-holonomy manifolds with emphasis on the Joyce-Karigiannis construction which will feature throughout this paper.
For reasons of brevity our description will necessarily be sketchy with no analytic details at all on the existence results,
but we encourage the reader to consult the original papers for more details.

The first examples of compact manifolds with $G_2$-holonomy are due to Joyce \cite{joyce1996compact1,joyce1996compact2}. These were constructed by a generalised Kummer construction, where one begins with a finite quotient of a 7-torus, $T^7 /\Gamma$ which is a singular $G_2$-orbifold. Then, for suitable choices of $\Gamma$, one can remove the singular set and glue in special holonomy model spaces to produce a smooth 7-manifold, which in favourable circumstances, will have a $G_2$-structure which is approximately close to being $G_2$-holonomy. For such $G_2$-structures, Joyce's main existence theorem asserts that one can perturb this $G_2$-structure to a genuinely $G_2$-holonomy structure. We will meet some explicit examples in section four.

Another construction is the twisted connected sum construction of \cite{kovalev2003twisted,CHNP,Nordstrom} in which one glues together a pair of asymptotically cylindrical Calabi-Yau three folds times a circle in a specific way, proves the existence of an approximately $G_2$-holonomy structure and again one applies Joyce's existence theorem.

More recently, Joyce and Karigiannis constructed $G_2$-holonomy manifolds by resolving codimension four orbifold singularities. These are the focus of this paper. The starting point is $(X_0, \varphi_0)$, a compact $G_2$-holonomy orbifold with torsion free $G_2$-structure $\varphi_0$. Further suppose that the orbifold singularities occur in codimension four along a connected 3-manifold $L$. Then the singularities must be of $ADE$ type, i.e. the fibers of the normal bundle to $L$ will be of the form 
$\mathbb{R}^4 /\Gamma_{ADE}$ with $\Gamma_{ADE}$ a finite subgroup of $\op{SU}(2)$ acting irreducibly on ${\mathbb R^4}$. Joyce-Karigiannis restrict to the simplest case when $\Gamma=\mathbb{Z}_{2}$. They show that under certain conditions which we describe shortly the orbifold singularities of $(X_0, \varphi_0)$ can be desingularised, by excising a neighbourhood of $L$ and gluing in a certain family of Eguchi-Hanson 4-manifolds, $M_{EH}=T^*S^2$ parametrised by $L$. A key assumption is that $L$ admits a nowhere vanishing harmonic 1-form, $\alpha_c$, with respect to the induced metric on $L$. The volume of the sphere at the origin of $T^*S^2$ is controlled by the norm of $\alpha_c$ times an overall scale, $t$ i.e. Vol$(S^2)= \pi t |\alpha_c|$
This produces a smooth 7-manifold, $X_c$ which they prove has metrics of $G_2$-holonomy. They also consider a ${\mathbb Z_2}$-twisted version of the construction which produces a different 7-manifold, $X_h$. In these constructions, the model space $M_{EH} \times L$ does not have a known exactly $G_2$-holonomy metric. This makes this construction different to those described above, since it is not based on gluing together model metrics. However, remarkably the authors are able to prove that suitable cancellations  occur allowing for the existence of an approximately $G_2$-holonomy structure on the compact 7-manifold such that Joyce's existence result can again be applied. This proves that the 7-manifolds $X_c$ and $X_h$ admit metrics with $G_2$-holonomy. We will need to describe some aspects of the topology of these manifolds.

In the first case a small neighbourhood of the singular set is removed and replaced by $M_{EH} \times L$. This gluing procedure increases both the second and third Betti numbers, in the sense that $b^i(X_c)=b^i(X_0)+b^{i-2}(L)$ for $i=2,3$. The induced metric on $M_{EH}$ admits a harmonic 2-form, $\beta$, essentially the Poincare dual of the zero-section of $T^*{S^2}$ and $\beta$ extends to a harmonic form in $X_c$. The wedge product of $\beta$ with $\alpha_c$ is a harmonic 3-form on $X_c$. Furthermore, if $\gamma$ is any other harmonic 1-form on $L$, then $\alpha_c +\epsilon \gamma$ will also be nowhere vanishing for small $\epsilon$. This explains the Betti numbers of $X_c$. In terms of homology, the Poincare dual of $\beta$ in $X_c$ is a 5-cycle of topology $S^2 \times L$, where $S^2$ can be thought of as the zero-section of $T^*S^2$. The dual of $\alpha_c\wedge\beta$ is a 4-cycle with topology $S^2\times\Sigma$, with $\Sigma$ being the Poincare dual of $\alpha_c$ in $L$.

In the ${\mathbb Z_2}$-twisted case, the singular set is replaced by $(M_{EH}\times\hat{L})/{\mathbb Z_2}$ where the ${\mathbb Z_2}$ acts non-trivially on $M_{EH}$. This is a non-trivial fibration over $L = \hat{L}/{\mathbb Z_2}$ with $M_{EH}$ fibres. A key fact is that $\beta$
is odd under this action and hence $X_h$ does not inherit any additional harmonic 2-forms from the gluing and $b^2(X_h)=b^2(X_0)$. The 3-form
$\alpha_h\wedge\beta$, however, is ${\mathbb Z_2}$-invariant and becomes a harmonic 3-form on $X_h$ and hence $b^3(X_h)=b^3(X_0)+b^1(L,{\mathbb Z_2})$, where the last term is the number of ${\mathbb Z_2}$-twisted harmonic 1-forms on $L$. This is equal to the number of independent nowhere vanishing harmonic 1-forms on $\hat{L}$ which are ${\mathbb Z_2}$-odd.
The Poincare dual of this harmonic 3-form is topologically of the form $(S^2 \times \hat{\Sigma})/{\mathbb Z_2}$, where $S^2$
is the zero section in $M_{EH}$ and $\hat{\Sigma}$ is the Poincare dual of $\alpha_h$ in $\hat{L}$.
We note in passing that these and other ${\mathbb Z_2}$-twisted harmonic fields have a variety of applications in mathematical gauge theories in various dimensions \cite{Taubes2014zero, doan2021existence, Taubes2020}

The reason that the harmonic 1-form is assumed to have no zeroes is because the volumes of the spheres in the Eguchi-Hanson spaces is directly proportional to the norm of $\alpha_{c,h}$. If $\alpha_{c,h}$ were allowed to have a zero the spheres would collapse to a point there and the total space would develop an additional singularity. Unfortunately, having control over the metric and curvature tensor in this more general situation is rather difficult, hence the assumption that $\alpha_{c,h}$ has no zeroes. Physically, as we discuss in the next section, one actually expects the existence of light degrees of freedom, in fact chiral fermions, when $\alpha_{c,h}$ has isolated zeroes.

\section{Interpretation in $M$-theory}

$M$-theory compactified on a manifold of $G_2$-holonomy $(X,\varphi)$ gives rise semi-classically to a four dimensional supergravity theory with $b^2(X)$ $\op{U}(1)$ vector multiplets, $b^3(X)$ neutral chiral multiplets, $\Phi_i$, and four supercharges ($i=1$....$b^3(X)$). The complex scalar fields, $\phi_j=t_j + is_j$, in the chiral multiplets contain axions, $t_j$, from harmonic modes of the 3-form field $C$ and the moduli, $s_j$, of the $G_2$-holonomy metric which appear as harmonic deformations of $\varphi$.

Additional, physically relevant light particles can arise if $X$ has special kinds of singularities.
Non-abelian gauge fields of type $ADE$ arise if $X$ contains codimension four orbifold singularities of type $ADE$ \cite{bsa2, acharya1999m}. Chiral fermions charged under such gauge symmetries will arise from additional, special kinds of conical codimension seven singularities \cite{ew2}.

In the Joyce-Karigiannis construction, we consider a $G_2$-orbifold with a codimension four $A_1$-singularity along a 3-manifold $L \subset X$, hence our story begins with an $\op{SU}(2)$ gauge theory on $L\times {\mathbb R^{3,1}}$ with the latter factor being our four dimensional spacetime. Since $G_2$-holonomy preserves supersymmetry, this $\op{SU}(2)$ gauge theory is supersymmetric. By integrating over $L$ and neglecting massive modes our goal is to provide a complete description of the low energy dynamics of this $\op{SU}(2)$ gauge theory in the form of a four-dimensional effective field theory.

These $M$-theory backgrounds have been analysed previously, beginning in \cite{acharya1999m}, and later in \cite{acharya2002moduli,pantev2011hitchin,Barbosa:2019bgh,Braun:2018vhk}. To briefly summarise the analysis, one is considering 7d $\op{SU}(2)$ supersymmetric Yang-Mills theory compactified on $L$. 
The 7d theory in flat space is known to have three scalar fields $\Vec{\phi}$, each in the adjoint representation of $\op{SU}(2)$. When compactified on $L$ in a $G_2$-orbifold the three fields become the components of a 1-form field $B$, again in the adjoint of $\op{SU}(2)$. We thus have a Yang-Mills gauge field $A$ and a 1-form Higgs field $B$ as the bosonic fields on $L$. These fields naturally pair up into a complex gauge field ${\cal{A}} = A + i B$ and the conditions on $\cal{A}$ which minimise the potential whilst preserving supersymmetry is that $\cal{A}$ is a harmonic flat connection on $L$ \cite{acharya2002moduli, pantev2011hitchin}. The space of classical vacua of the low energy effective theory is therefore the space of flat complex $\op{SU}(2)$ connections on $L$. In general, this space will have distinct disconnected components.
As we will show, the distinct components naturally correspond to the topologically distinct desingularisations of $X_0$ constructed in \cite{joyce2021new}. Previous analyses have focused on the flat connections continuously connected to the identity.

\subsection{The Coulomb Phase}
 
The identity connected component of the space of flat $\op{SU}(2)$ connections is $b^1(L)$-dimensional. Once complexified by $B$ we obtain $b^1(L)$ massless chiral multiplets in the four-dimensional effective theory. These naturally match up with the $b^1(L)$ moduli of $X_c$ which desingularise the orbifold $X_0$, as reviewed above. 

A crucial fact about the Joyce-Karigiannis theorem is the assumption that the harmonic 1-form must be nowhere vanishing; whereas in the physical  analysis, the nowhere zero condition is generally not required. The harmonic 1-form $\alpha_c$ which appears in the Joyce-Karigiannis theorem is identified with $B$ in the direction of the Cartan subalgebra of $\op{SU}(2)$ and further can be identified with the volume and complex structure of the $S^2$ in the centre of $T^*S^2$ as it varies over $L$. If $\alpha_c$ had a zero at a point $p$, $B$ would vanish and hence, $\op{SU}(2)$ gauge symmetry is restored at $p$. 
Geometrically, the norm of $\alpha_c$ controls the size of the two-sphere of $M_{EH}$, hence, away from $p$ the glued in $M_{EH}$ spaces are smooth. But at $p$ the $M_{EH}$ degenerates to an orbifold. At this point we expect that $X$ itself develops a further singularity. In fact, the cone over $\mathbb{CP}^3$ i.e. $\mathbb{R}^{+}\times \mathbb{CP}^3$ is, topologically, a 3-dimensional family of Eguchi-Hanson spaces which at the origin degenerate to $\mathbb{R}^4/\mathbb{Z}_{2}$ and this was precisely the description given in \cite{ew2}, where this additional singularity is interpreted as giving rise to a chiral fermion charged under the $\op{U}(1)$ gauge symmetry. One would certainly like to have a better understanding of what the zeroes of harmonic 1-forms on $L$ imply physically and for the would-be $G_2$-holonomy space $(X,\varphi)$.

In any case, when $\alpha_c$ has no zeroes, at a generic point in moduli space, $\op{SU}(2)$ is broken to its maximal torus and hence we refer to this branch of vacua as the Coulomb branch (hence the subscript on $\alpha_c$). Hence, the low energy description is simply an $\mathcal{N}=1$ supersymmetric $\op{SU}(2)$ gauge theory with $b^1(L)$ adjoint chiral multiplets, as originally found in \cite{acharya1999m}.

\subsection{The Higgs Phase}

Interpretation of the dynamics in $M$-theory on $X_h$ is one of the main results. We will see that in this case there is a non-identity connected component of the space of flat $\op{SU}(2)$-connections, which naturally corresponds to the moduli of $X_h$.
In this case we have $L=\hat{L}/\mathbb{Z}_{2}$ and $\hat{L}$
has a $\mathbb{Z}_{2}$-odd harmonic 1-form, $\alpha_h$, with no zeroes. The existence of $\alpha_h$ implies that $b_1(\hat{L})$ is at least one and that $\pi_1 (\hat{L})$ contains an element $g_{\alpha}$ of infinite order. Furthermore, this element is $\mathbb{Z}_{2}$ odd, hence, if $g_\beta$ gives the order two action on $\hat{L}$,
\be
g_\beta g_\alpha g_\beta^{-1} = g_{\alpha}^{-1} 
\ee
In general, although $g_\beta$ is of order two on $\hat{L}$ it need not be of order two on its universal cover. Hence
\be
g_\beta^2 = g_{\gamma} 
\ee
for some other element $g_\gamma$.
Vacua of the 7d $\op{SU}(2)$ Yang-Mills theory on $L$ are given by specifying a flat $\op{SU}(2)$ connection on $L$. Modulo conjugation, these are just given by  set of matrices in $\op{SU}(2)$, satisfying the relations of $\pi_1(L)$. In particular, we would like to satisfy the above two relations with $\op{SU}(2)$ matrices, $M_{\alpha}, M_{\beta}, M_{\gamma}$.

Without loss of generality, we can conjugate $M_{\alpha}$ into the maximal torus and hence take $M_{\alpha}$ to be diagonal.
The first relation then asserts that $M_{\beta}$ {\it permutes} the two eigenvalues of $M_{\alpha}$ and is thus in the Weyl group of $\op{SU}(2)$:
\begin{equation}
M_{\alpha} = \begin{pmatrix}
e^{i\theta_{3}}&0\\0&e^{-i\theta_{3}}
\end{pmatrix},\quad M_{\beta} = \begin{pmatrix}
0&1\\-1&0
\end{pmatrix},\quad M_{\gamma} = -\mathbb{1}
\end{equation}

We thus see that we have a one-dimensional space of vacua, which is in keeping with the one dimensional space of $G_2$-manifolds, $X_h$,  constructed by Joyce and Karigiannis. The fact that the Weyl group plays a key role is essential and was already anticipated by Joyce \cite{joyce1998topology}. Also, though obvious, we note that these flat connections cannot be continuously deformed to the identity.

Since the subgroup of $\op{SU}(2)$ generated by $M_{\alpha}$ and $M_{\beta}$ break $\op{SU}(2)$ to its centre, $\mathbb{Z}_{2}$, this tells us that at generic points in its space of vacua, the gauge group of the low energy theory is broken to $\mathbb{Z}_{2}$, however, since there are no fields in the 7d theory which are charged under the centre, the classical low energy theory has the gauge group effectively broken completely.
As we will see, this component of the moduli space corresponds to a Higgs phase in the effective four dimensional theory, and hence the subscript on $\alpha_h$.

The key is to note that at the origin of the moduli space, $M_{\alpha}=\mathbb{1}$ and that the $\op{SO}(2)$ subgroup of $\op{SU}(2)$ consisting of real matrices remains unbroken by $M_{\beta}$. This tells us that the gauge group of the four dimensional theory is $\op{SO}(2)$. 
There must therefore also be supersymmetric Higgs fields whose vacuum values break $\op{SO}(2)$ completely. The proposal is that there are precisely two complex, chiral superfields, $\Phi_{1,2}$ transforming in the fundamental representation of $\op{SO}(2)$. This Higgs doublet contains four bosonic degrees of freedom of which one becomes the longitudinal component of the now massive gauge boson. Another is the Higgs boson itself and these two massive degrees of freedom comprise the degrees of freedom of a massive vector multiplet.   The remaining two degrees of freedom remain massless and give rise to the expected complex one-dimensional space of vacua arising from the Joyce-Karigiannis construction.

The Higgs doublet is naturally associated with the $\mathbb{Z}_{2}$-twisted harmonic 1-form $\alpha_h$ because at the origin of the moduli space the flat $\op{SU}(2)$ connection in the adjoint representation arises from a $\mathbb{Z}_{2}$-bundle via the natural inclusions $\mathbb{Z}_{2} \subset \op{SO}(3) \leftarrow \op{SU}(2)$. Hence, at the origin of the moduli space, the Yang-Mills Laplacian effectively reduces to the Laplacian acting on $\mathbb{Z}_{2}$-twisted 1-forms. Another way to obtain this result is that if we look at how the $\mathbb{Z}_{2}$ acts on the fields on $\hat{L}$ it is via the combined action of the geometric action together with the gauge transformation by $g_\beta$ and it is the $\op{SO}(2)$ doublet of fields in the low energy theory which are $\mathbb{Z}_{2}$-invariant.

To summarise: The $\mathbb{Z}_{2}$-twisted Joyce-Karigiannis construction of $G_2$-manifolds $(X_h, \varphi)$ has a low energy description as a supersymmetric $\op{SO}(2)$ gauge theory with matter in the fundamental representation. The one-dimensional complexified moduli space of $G_2$-holonomy metrics correpsonds naturally to the Higgs branch of this gauge theory.

Note that, in general $L$ could have both ordinary harmonic as well as $\mathbb{Z}_{2}$-twisted harmonic 1-forms. In this case, there will clearly be both Coulomb and Higgs vacua arising from the same $G_2$-orbifold.

The picture developed above can clearly be generalised in two ways. First, it is clear that one can consider more general $ADE$ singularities beyond $\op{SU}(2)$. Second, one may also consider higher order twists beyond $\mathbb{Z}_{2}$. We will encounter both of these possibilities in what follows.

In the next section we introduce  some simple explicit local models which are exactly solvable and which allow us to consider any $ADE$ gauge group as well as higher order twists where we can prove that the gauge theory moduli space is the moduli space of $G_2$-holonomy metrics desingularising $X_0$.
Following that we will describe some compact $G_2$-manifolds which give rise to both Coulomb and Higgs branches classically.

\section{Explicit Local Models}\label{sec: 4.1}

The simplest models of 3-manifolds, $L$,  admitting nowhere vanishing $\mathbb{Z}_{2}$-twisted harmonic 1-forms are when $\hat{L} = \Sigma \times S^1$ with Riemannian product metric where the $\mathbb{Z}_{2}$ acts simultaneously as an orientation reversing isometry of both the $S^1$ and the compact Riemann surface $\Sigma$. Then, the standard harmonic 1-form on $S^1$ is nowhere vanishing and $\mathbb{Z}_{2}$-twisted in the quotient $L$. We can simplify this even further by considering $\Sigma=T^2$ with a flat metric i.e. we can take
$L$ to be a smooth $\mathbb{Z}_{2}$-quotient of the flat 3-torus. Since $L$ is oriented, the $\mathbb{Z}_{2}$ action is essentially unique.
If the coordinates of 3-torus are denoted as $y_{1,2,3}$ with periodicities chosen to be one, then the $\mathbb{Z}_{2}$ action may be defined as:

\begin{equation}
    (y^{1},y^{2},y^{3})\mapsto (-y^{1},-y^{2},y^{3}+1/2).
\end{equation} 

We see that $dy^1$ and $dy^2$ are both $\mathbb{Z}_{2}$-twisted harmonic 1-forms, whilst $dy^3$ is an ordinary harmonic 1-form. Therefore if this $L$ arises in the Joyce-Karigiannis construction, there will be a 2-parameter family of $G_2$-manifolds corresponding to a Higgs branch and a 1-parameter Coulomb branch\ft{Joyce explicitly constructs examples of compact $G_2$-manifolds by desingularising such singularities as we will describe later.}.  We will next explicitly construct local models of such singular $G_2$-spaces before exhibiting the corresponding moduli spaces of flat $\op{SU}(2)$ connections.

\subsection{A Simple Example}

We are thus looking for a local $G_2$-holonomy orbifold $X_0$ which fibers over $L$ with fibers $\mathbb{C}^{2}/\mathbb{Z}_{2}$. The simplest model arises by taking $L$ to be flat and, since the ambient metric is Ricci flat, the 7-orbifold itself will be locally a Riemannian product i.e.

\begin{equation}\label{eq:T3 times Ak mod Z2}
   X_o= \frac{(T^{3}\times {\mathbb{C}^{2}}/{\mathbb{Z}_{2}})}{\mathbb{Z}_{2}}
\end{equation}
with a flat metric. These locally flat models are {\it very special cases} of the Joyce-Karigiannis construction where the harmonic 1-form is constant along $L$.
The fibration over $L$ is however non-trivial as the requirement that the holonomy be contained in $G_2$ requires the $\mathbb{Z}_{2}$ which acts on $T^{3}$ to also act on $\mathbb{C}^{2}$.
In fact, we can choose complex coordinates  $(z_1, z_2)$ in which the $\mathbb{Z}_{2}$-action is 
$(z_1, z_2)\rightarrow (-z_1, z_2)$.  

The torsion-free $G_{2}$-structure on \eqref{eq:T3 times Ak mod Z2} is,
\begin{equation}
    \varphi = dy^{1}dy^{2}dy^{3} + d\vec{y}\cdot\vec{\omega},
\end{equation}
where $\omega_{i}$ are the K\"ahler 2-forms on $\mathbb{C}^{2}/\mathbb{Z}_{k}$ defining the flat hyperK\"ahler structure. Then the metric is,
\begin{equation}
    g = d\vec{y}^{2} + h
\end{equation}
where $h$ is the Euclidean metric on $\mathbb{C}^{2}/\mathbb{Z}_{k}$. 

The $G_2$-orbifold $(X_0,\varphi)$ admits two topologically distinct smooth desingularisations $(X_c, \varphi_c)$ and $(X_h, \varphi_h)$ of the form

\begin{equation}\label{eq:T3 times EH mod Z2}
    \frac{T^{3}\times M_{EH}}{\mathbb{Z}^a_{2}},\;\;\; a=c,h
\end{equation}
with Ricci flat metrics

\begin{equation}
    g_a = d\vec{y}^{2} + h_{EH}
\end{equation}
where $h_{EH}$ is the family of hyper-K\"ahler Eguchi-Hanson metrics on $T^* S^2$. Both of these are just free $\mathbb{Z}_{2}$ quotients of $M_{EH} \times T^3$, differing by the action of the involution: as described in section two, in $X_c$, $H_2(M_{EH})$ is preserved by the $\mathbb{Z}_{2}$, whereas in $X_h$ it is odd.
Note that the holonomy group of both of the metrics $g_a$ is $\op{SU}(2)\ltimes \mathbb{Z}_{2} $ and that this group is a subgroup of $\op{SU}(3) \subset G_2$. Hence these special local models actually preserve $\mathcal{N}=2$ supersymmetry in four dimensions\footnote{We will describe a genuinely $\mathcal{N}=1$ supersymmetric example at the end of this subsection.}.

$X_c$ and $X_h$ are topologically distinct since $b^2(X_c)=b^3(X_c)=1$, whilst $b^2(X_h)=0$ and $b^3(X_h)=2$. 
Hence, whilst $X_c$ has a one-dimensional moduli space of $G_2$-metrics, the moduli space of $X_h$ is two-dimensional.
The topology of the compact 4-cycles which arise from the two distinct desingularisations are $S^2 \times T^2/{\mathbb Z_2}$ on the Coulomb branch and $(S^2 \times T^2)/\mathbb{Z}_{2}$ in the Higgs case. The latter may be regarded as the non-trivial $S^2$-bundle over the Klein bottle or as a particular $T^2$-fibration over $\mathbb{RP}^{2}$. All of these cycles have calibrated (co-associative), i.e. supersymmetric, representatives.

\subsubsection{Flat Connections on $L$}\label{sec:4.1.1}
Since this example is so explicit we can also be explicit about the gauge theory interpretation. The low energy field theory descriptions in four dimensions will be given by gauge theories whose moduli spaces of vacua are the moduli spaces of flat connections on $L$. These models were first considered in the physics literature in \cite{acharya1999m} and the moduli spaces were considered by Barbosa in his PhD thesis \cite{barbosa2019thesis}.

The fundamental group of $L$ in this case can be presented with four generators: three translations representing the fundamental group of $T^3$ and another representing the $\mathbb{Z}_{2}$ quotient. The explicit relations may be presented as:

\begin{equation}
\left\langle g_{1},\;g_{2},\;g_{3},\;g_{\beta}\mid g_{i}g_{j}=g_{j}g_{i}\quad i=1,2,3,\quad g_{\beta}^{2}=g_{3},\quad g_{\beta} g_{3}g_{\beta}^{-1}=g_{3}, \quad g_{\beta} g_{1,2}g_{\beta}^{-1}=g_{1,2}^{-1}\right\rangle
\end{equation}

There are several components to the moduli space of flat $\op{SU}(2)$ connections on $L$. First there are actually four, one dimensional Coulomb branches: 
\begin{equation}
g_{1}\mapsto\pm\begin{pmatrix}
1&0\\0&1
\end{pmatrix},\; g_{2}\mapsto\pm\begin{pmatrix}
1&0\\0&1
\end{pmatrix},\; g_{3}\mapsto\begin{pmatrix}
e^{i\theta_{3}}&0\\0&e^{-i\theta_{3}}
\end{pmatrix},\; g_{\beta}\mapsto\begin{pmatrix}
e^{i\theta_{3}/2}&0\\0&e^{-i\theta_{3}/2}
\end{pmatrix}
\end{equation}
Note that, at the origin ($\theta_{3}=0$), the centraliser of the solution in $\op{SU}(2)$ is the whole group and away from the origin we break down to the maximal torus $\op{U}(1)$. 

Noticing that the last two group relations are exactly as described in the previous section, we learn that there is also a Higgs branch:
\begin{equation}\label{Higgs solution}
g_{1}\mapsto\begin{pmatrix}
e^{i\theta_{1}}&0\\0&e^{-i\theta_{1}}
\end{pmatrix},\; g_{2}\mapsto\begin{pmatrix}
e^{i\theta_{2}}&0\\0&e^{-i\theta_{2}}
\end{pmatrix},\; g_{3}\mapsto\begin{pmatrix}
-1&0\\0&-1
\end{pmatrix},\; g_{\beta}\mapsto\begin{pmatrix}
0&1\\-1&0
\end{pmatrix}
\end{equation}

These are all the flat $\op{SU}(2)$-connections and this result agrees with \cite{barbosa2019harmonic,barbosa2019thesis}.
The Coulomb branch solutions have one parameter $\theta_3$, corresponding to $b^3(X_c)=1$, whilst the Higgs branch has two parameters as expected by $b^3(X_h)=2$ in perfect agreement with the moduli space of Ricci flat, special holonomy metrics described above.

\subsubsection{Field Theory Interpretation}
The four dimensional field theory description is now straightforward:
The classical dynamics on each of the four Coulomb branches is just pure $\mathcal{N}=2$ $\op{SU}(2)$
Yang-Mills theory. The Higgs branch description is given by $\mathcal{N}=2$ supersymmetric $\op{SO}(2)$ Yang-Mills with a hypermultiplet in the fundamental representation. At first sight it seems that there is a discrepancy in the comparison to the moduli space of $M$-theory in the Coulomb phase, since there is only one $G_2$-manifold $X_c = \frac{T^{3}\times M_{EH}}{\mathbb{Z}^c_{2}}$ but four Coulomb branches. However, there are four spaces of vacua arising from $X_c$ differing by the expectation values of the $C$-field. Flat $C$-fields on $X_c$ are classified by $H^3(X_c, \op{U}(1))$ and this is given by $\mathbb{Z}_{2} \times \mathbb{Z}_{2} \times \op{U}(1)$ and there are therefore four discrete families of one-dimensional flat $C$-field backgrounds. One may also think of these as the $\mathbb{Z}_{2}^c$-invariant harmonic $C$-fields on ${T^{3}\times M_{EH}}$.

These locally flat explicit models clearly have natural generalisations to $G_2$-orbifolds of the form
\begin{equation}
   X_o(\Gamma_{ADE}, K) = \frac{\mathbb{C}^{2}/\Gamma_{ADE} \times T^3}{K}
\end{equation}
where $K$ is a finite group acting freely on $T^{3}$, preserving orientation. The existence of these examples demonstrate that one can consider more general gauge groups as well as $K$-twisted harmonic 1-forms.
In these examples, $T^{3}/K$ is an orientable, compact Bieberbach 3-manifold, hence there are only six possibilities: $K=
1,\mathbb{Z}_{2},\mathbb{Z}_{3},\mathbb{Z}_{4},\mathbb{Z}_{6},\mathbb{Z}_{2} \times \mathbb{Z}_{2}$. The key to understanding the moduli spaces of Ricci flat metrics on these spaces reduces essentially to classifying the actions of $K$ on $\mathbb{C}^{2}/\Gamma_{ADE}$ which lift to actions on its hyperK\"ahler desingularisations, $\widetilde{\mathbb{C}^{2}/\Gamma_{ADE}}$.
This amounts to classifying actions of $K$ on 
$(\widetilde{\mathbb{C}^{2}/\Gamma_{ADE}} \times T^{3})$ which preserve the $G_2$-structure, giving rise to Ricci flat metrics with holonomy group $\op{SU}(2) \ltimes K$ on the smooth 7-manifolds $(\widetilde{\mathbb{C}^{2}/\Gamma_{ADE}} \times T^{3})/K$. These metrics preserve four dimensional $\mathcal{N}=2$ supersymmetry for 
$K=
1,\mathbb{Z}_{2},\mathbb{Z}_{3},\mathbb{Z}_{4},\mathbb{Z}_{6}$, but the case $K=\mathbb{Z}_{2} \times \mathbb{Z}_{2}$ has $\mathcal{N}=1$ supersymmetry. This latter example was considered in \cite{acharya1999m} and the moduli spaces of flat $\op{SU}(2)$-connections in \cite{barbosa2019harmonic}. We will be able to describe all components of the moduli space of Ricci flat metrics and compare that with the space of flat connections and then describe the low energy dynamics of the effective four dimensional field theory associated to each branch.

\subsection{$\mathcal{N}=1$ Supersymmetric Example}.

The 3-manifold in this example is $L = T^3/\mathbb{Z}_{2} \times \mathbb{Z}_{2}$. The action of $K$ on the coordinates of $\mathbb{C}^{2}/\mathbb{Z}_{2} \times T^3$ is given by:
\begin{align}
    g_{\beta_{1}}: &(z^1, z^2, y^{1},y^{2},y^{3})\mapsto (-z^1, z^2, y^{1}+1/2,-y^{2},-y^{3})\\
    g_{\beta_{2}}: &({z}^1, {z}^2, y^{1},y^{2},y^{3})\mapsto (\bar{z}^1, \bar{z}^2, -y^{1},y^{2}+1/2,-y^{3}+1/2)\\
    g_{\beta_{3}}: &({z}^1, {z}^2, y^{1},y^{2},y^{3})\mapsto (-\bar{z}^1, \bar{z}^2, - y^{1}+1/2,-y^{2}+1/2,y^{3}+1/2)
\end{align}

The fundamental group of $L$ can thus be described as having six generators $g_{1,2,3}$, $g_{\beta_{1}}$, $g_{\beta_{2}}$, $g_{\beta_{3}}$ with the relations

\begin{align}
\langle g_{1},\;g_{2},\;g_{3},\;g_{\beta_{1}},\;g_{\beta_{2}},\;g_{\beta_{3}}\mid & g_{i}g_{j}=g_{j}g_{i}, \quad g_{\beta_{i}}^{2}=g_{i},\quad g_{\beta_{i}} g_{i}g_{\beta_{i}}^{-1}=g_{i}, \quad i=1,2,3,\nonumber\\
& g_{\beta_{i}} g_{j,k}g_{\beta_{i}}^{-1}=g_{j,k}^{-1},\quad i\neq j \neq k, \quad g_{\beta_{3}} g_{\beta_{2}}g_{\beta_{1}}=g_1 g_3  \rangle
\end{align}
Here $g_{\beta_{3}}=g_{1} g_{\beta_{2}} g_{\beta_{1}}$.

\subsubsection{Moduli space of Ricci flat metrics}

In order to describe the possible smooth Ricci flat manifolds $(M_{EH} \times T^{3})/K$ we first have to describe how the $\mathbb{Z}_{2}$-symmetries $g_{\beta_{1,2,3}}$ act on $M_{EH}$. Since $H_2(M_{EH})=\mathbb{Z}$, generated by the $S^2$ in the centre, each $g_{\beta_{i}}$ can either preserve or reverse the orientation of all classes in $H_2(M_{EH})$.
The action on homology is therefore specified by a sign for each of the three involutions. 
There are then three smooth possibilities according to how the $g_{\beta_{i}}$ act on the $S^2$ at the centre of $M_{EH}$. These are given by the choices
$(i): (+,-,-)$, $(ii): (-,+,-)$ and $(iii): (-,-,+)$, where, for example $(+,-,-)$ means that $g_{\beta_{1}}$ preserves $H_2(M_{EH})$ whilst $g_{\beta_{2}}$ and $g_{\beta_{3}}$ act as minus the identity. Since there is no $(+,+,+)$ case there are no compact 2-cycles or 5-cycles in $(M_{EH} \times T^3)/K$ or equivalently no compactly supported harmonic 2-forms. In $M$-theory these would have given rise to $\op{U}(1)$ gauge fields in four dimensions and hence a Coulomb branch. Thus there is no continuous Coulomb branch of vacua. 
In case $(i)$ we see that there are compact 4-cycles in $(M_{EH} \times T^{3})/K$ which are Poincare dual to $\beta \wedge dy^1$. In case $(ii)$ instead, it is $\beta \wedge dy^2$ which is invariant and in case $(iii)$ the Poincare dual of $\beta \wedge dy^3$ is an invariant compact 4-cycle. The proof that these are the only possibilities will be given section 4.4. The space of Ricci flat metrics resolving the orbifold singularities thus has three one-dimensional components, giving a space of vacua which is $\mathbb{C} \cup \mathbb{C} \cup \mathbb{C}$.

We will now examine the moduli space of flat $\op{SU}(2)$-connections on $L$ and see that it matches nicely with this description.

\subsubsection{Flat Connections on $L$}\label{sec:4.2.2}

\bigskip

{\it Coulomb Branches.}

\bigskip

First note that, since none of the $dy^{i}$ are $K$-invariant, $b^{1}(L)=0$, so there is no continuous moduli space of Coulomb vacua. 
There are non-trivial, discrete, Abelian connections however. Note that the Abelianisation of $\pi_{1}(L)$ is $H_{1}(L, \mathbb{Z})=\mathbb{Z}_{4} \times \mathbb{Z}_{4}$. This can be seen to be generated by $g_{\beta_{1}}$ and $g_{\beta_{2}}$ with the relations that both are order four. Thus the corresponding sixteen flat $\op{SU}(2)$ connections on the Coulomb branch can be obtained by choosing $g_{\beta_{1}}$ and $g_{\beta_{2}}$ independently from the four diagonal matrices in the $\mathbb{Z}_{4}$ subgroup of the maximal torus.
\bigskip

\noindent{\it Higgs Branches.}

\bigskip

From the relations of the fundamental group as we have presented them, the strategy to finding flat connections should be clear. One chooses a flat connection in the Weyl group of $\op{SU}(2)$ for each of the three non-trivial elements of $K$. The remaining elements of the group are diagonal. This will give rise to three components to the space of flat connections, beyond the Coulomb branch above. The three components are, for $i\neq j \neq k$,

\begin{align}\label{eq:higgs solution z2xz2}
g_{i}\mapsto\begin{pmatrix}
e^{i\theta_{i}}&0\\0&e^{-i\theta_{i}}
\end{pmatrix},\; g_{j}&\mapsto\begin{pmatrix}
-1&0\\0&-1
\end{pmatrix},\; g_{k}\mapsto\begin{pmatrix}
-1&0\\0&-1
\end{pmatrix},\; \\ g_{\beta_i}\mapsto\begin{pmatrix}
e^{i\theta_{i}/2}&0\\0&e^{-i\theta_{i}/2}
\end{pmatrix},\; g_{\beta_{j}}&\mapsto\begin{pmatrix}
0&1\\-1&0
\end{pmatrix},\; g_{\beta_{k}}\mapsto\begin{pmatrix} 
0&e^{i\theta_{i}/2}\\-e^{-i\theta_{i}/2}&0
\end{pmatrix}
\end{align}

The interpretation in the the low energy effective theory of each branch is clear: an $\mathcal{N}=1$ supersymmetric $\op{SO}(2)$ gauge theory with chiral multiplets in the fundamental representation and a flat direction in the space of vacua along which the gauge group is spontaneously broken.
The fact that there are three one dimensional Higgs branches matches the three one-dimensional components to the moduli space of Ricci flat metrics found in the previous subsection.

\subsection{Moduli Space of Ricci Flat Metrics}

In this subsection we describe the space of Ricci flat metrics with holonomy group $\op{SU}(2)\rtimes K$ on the smooth families of 7-manifolds $X_a(\Gamma_{ADE}, K):=(\widetilde{\mathbb{C}^{2}/\Gamma_{ADE}} \times T^{3})/K$ which desingularise the flat orbifolds 
\begin{equation}
   X_o(\Gamma_{ADE}, K) = \frac{\mathbb{C}^{2}/\Gamma_{ADE} \times T^{3}}{K}
\end{equation}
and the parameter $a$ collectively denotes the Coulomb or Higgs branch parameters of the family.

Since $X_a(\Gamma_{ADE}, K)$ is simply a free quotient of $\widetilde{\mathbb{C}^{2}/\Gamma_{ADE}} \times T^{3}$, the moduli space of Ricci flat metrics is given by the subspace of Ricci flat metrics on $\widetilde{\mathbb{C}^{2}/\Gamma_{ADE}}$ which admit the appropriate action of $K$.
We will first describe the answer to the problem for  $X_a(\Gamma_{A_{n}}, K)$ where we can use the explicit form of the metrics given by Gibbons and Hawking \cite{GibbonsHawking}. This then provides enough insight to solve the problem completely in the general case.

\subsubsection{Multi-Centre Gibbons-Hawking Space}\label{sec:GH space}

An $A_{n-1}$ singularity is the singularity at the origin of $\mathbb{C}^2/\mathbb{Z}_{n}$, where $\mathbb{Z}_{n}$ acts as a subgroup of $\op{SU}(2)$. The spaces $\mathbb{C}^{2}/\Gamma$ (where $\Gamma$ is a finite subgroup of $\op{SU}(2)$) all have topologically unique crepant resolutions , $\widetilde{\mathbb{C}^{2}/\Gamma}$
which are hyperK\"ahler and Asymptotically Locally Euclidean (ALE) \cite{Kronheimer}. 

For the $A_{n-1}$ singularities, the corresponding ALE spaces are the $n$-centre Gibbons-Hawking spaces $M^{(n)}_{GH}$ with explicit metrics
given by
\begin{align}
ds^{2} &=g_{GH}= V(\vec{x})\;d\vec{x}\cdot d\vec{x} + V(\vec{x})^{-1}(dt+A_{i}\; dx^{i})^{2}\\
V(\vec{x}) &= \sum_{\gamma=1}^{n}\frac{1}{|\vec{x}-\vec{a}_{\gamma}|}\\
\vec{\nabla}\times\vec{A} &= \vec{\nabla}V(\vec{x})\quad\text{or equivalently}\quad \ast d A = dV.\label{eq:potential and connection condition}
\end{align}
where $\vec{x}\in \mathbb{R}^{3}$ and $t\in S^{1}$. There are 3$n$-parameters which appear as $n$ 3-vectors, $\vec{a}_{\gamma}$ and are the centres of the harmonic functions appearing in the potential $V(\vec{x})$. By a choice of coordinates we can assume that $\sum_\gamma \vec{a}_\gamma = 0$.

There is a triplet of complex structures, $(I,J,K)$, given by,
\begin{equation}
Idx^{1} = V(\vec{x})^{-1}(dt+\vec{A}\cdot d\vec{x}),\qquad Idx^{2} = dx^{3}
\end{equation}
with $J$ and $K$ given by cyclically permuting $dx^{1,2,3}$. The hyper-K\"ahler forms are determined from the metric via, $\omega_{I}(\cdot,\cdot) = g(I\cdot,\cdot)$ and similarly for $J$ and $K$. They are given by,
\begin{equation}\label{eq:GH HK forms}
\omega_{I}\equiv \omega_{1} = (dt+A_{i}\; d x^{i})\wedge dx^{1} + V(\vec{x})\;dx^{2}\wedge dx^{3}
\end{equation}
with $\omega_{J,K}$  obtained by cyclic permutations of $dx^{1,2,3}$.

The cohomology group $H^{2}\left(M_{GH}^{(n)},\mathbb{Z}\right)$ is spanned by the $L_{2}$-normalisable 2-forms \cite{Hausel:2002xg} which are the Poincar\'e duals of the 2-spheres arising from line segments in $\mathbb{R}^{3}$ connecting adjacent centres. Denote the 2-form associated to the centres $a_{\gamma}$ and $a_{\gamma+1}$ by $\Gamma_{\gamma}$. We can write these explicitly, first define,
\begin{equation}
    V_{i} \equiv \frac{1}{\lvert \vec{x}-\vec{a}_{\lambda}\rvert},\quad V \equiv \sum_{\lambda=1}^{n}V_{\lambda}
\end{equation}
then define the basis of anti-self-dual 2-forms,
\begin{equation}
    \Sigma^{a} = e^{a}\,e^{4}-\frac{1}{2}\,\varepsilon^{a}{}_{bc}\,e^{b}\,e^{c},\quad e^{1,2,3} = V^{1/2}dx^{1,2,3},\quad e^{4} = V^{-1/2}(dt+A_{i}\,dx^{i})
\end{equation}
To each centre we can associate a 2-form $\Omega_{\gamma} \equiv -\partial_{a}\left(V_{\gamma}/V\right)\Sigma^{a}$ and define the following 2-form,
\begin{equation}
    \Gamma_{\gamma} = -\frac{1}{4\pi}(\Omega_{\gamma}-\Omega_{\gamma+1}),\quad \gamma = 1,...,n-1
\end{equation}
which is anti-self-dual and $L_{2}$-normalisable. Further, one can check that,
\begin{equation}
    \int_{M_{GH}^{(n)}} \Gamma_{i}\wedge  \Gamma_{j}
\end{equation}
is minus the $A_{n-1}$ Cartan matrix using the fact that \cite{Ruback:1986ag,Sen:1997js},
\begin{equation}
    \int_{M_{GH}^{(n)}}\Omega_{i}\wedge \Omega_{j} = -16\pi^{2}\delta_{ij}
\end{equation}
Having an explicit expression for $\Gamma_{i}$ makes it easy to see how many $K$-invariant and $K$-twisted 2-forms we have on $X_a(\Gamma_{ADE}, K)$.

\bigskip

\subsubsection{Gibbons-Hawking Moduli space and Flat Connections on $T^3$}\label{sec:GH mod space and flat conn on T3}.

\bigskip

Here we explicitly demonstrate the relationship between the moduli space of flat $SL(n, \mathbb{C})$ connections on a 3-torus to the Gibbons-Hawking moduli space. We will show that the $M$-theory moduli space is isomorphic to the space of flat connections.

The $n$ centres, $\vec{a}_{\gamma}$ of the harmonic potential $V$ are 3n-free parameters whose sum is fixed. 
Hence the moduli space of Ricci flat metrics is given by $(\mathbb{R}^3)^{n-1} /S_{n}$, where we factor out by permutations of the centres, which acts as the Weyl group of $\op{SU}(n)$. In fact there is a close connection between this moduli space and the space of flat $\op{SU}(n)$ connections on $T^{3}$. A flat $\op{SU}(n)$  connection on $T^{3}$ is just given by three commuting elements of $\op{SU}(n)$ and, in fact, is given by any three elements of the maximal torus, $T(\op{SU}(n))$, modulo the action of the Weyl group \cite{FriedmanMorganWitten}. 
This space is $(T(\op{SU}(n)))^{3}/{S_n}$.
In $M$-theory this gets complexified to the space of $SL(n,\mathbb{C})$ connections which is essentially $(\mathbb{C}^{*3})^{n-1}/{S_n}$ and points in this space are just diagonal matrices of unit determinant, up to the Weyl group action. Denote these three diagonal matrices by $M_{a}$, $M_{b}$ and $M_{c}$ and their diagonal elements as $(\lambda_{a_{1}}, \lambda_{a_{2}}, \cdots ,\lambda_{a_{n}})$, $(\lambda_{b_1}, \lambda_{b_2}, \cdots ,\lambda_{b_{n}})$ and $(\lambda_{c_{1}}, \lambda_{c_{2}}, \cdots ,\lambda_{c_{n}})$. If we suggestively label the coordinates of a given centre by $(a_{\gamma} , b_{\gamma}, c_{\gamma})$, then
the absolute values of the diagonal entries of the matrices are the exponentials of the entries: $|\lambda_{a_{\gamma}}|=e^{a_{\gamma}}$, $|\lambda_{b_{\gamma}}|=e^{b_{\gamma}}$ and $|\lambda_{c_{\gamma}}|=e^{c_{\gamma}}$. This is the explicit relationship between the flat connections and the Ricci flat metrics. We can recover the full moduli space of flat $SL(n,\mathbb{C})$-connections by including the harmonic modes of the $C$-field in $M$-theory. In $M$-theory on $(\widetilde{\mathbb{C}^{2}/\mathbb{Z}_{n}}\times T^{3})$ with the metric $g_{GH} + h$, in addition to the Gibbons-Hawking moduli $(a_\gamma , b_\gamma, c_\gamma)$ we have the massless scalar fields arising from the harmonic modes of the $C$-field. These are given by $H^3(\widetilde{\mathbb{C}^{2}/\mathbb{Z}_n}\times T^{3}, \op{U}(1)) = (T(\op{SU}(n)))^3 \times S^1$, where the $S^1$ is the set of four-dimensional axion VEVs and will play no further role. However, this description is not complete since it does not take into account the action of the non-identity connected diffeomorphisms of $(\widetilde{\mathbb{C}^{2}/\mathbb{Z}_n}\times T^{3})$ given by the permutation group $S_n$. This group also acts as the Weyl group on $H^2(\mathbb{C}^{2}/\mathbb{Z}_{n})$ which induces the standard action of the Weyl group on the maximal torus $T(\op{SU}(n))$. Hence, for fixed axion vev, the moduli space of $C$-fields is given by $(T(\op{SU}(n)))^3/{S_n}$ and is precisely the moduli space of flat $\op{SU}(n)$ connections on $T^3$. This shows that, for fixed $T^3$ volume and axion VEV, that the moduli space of $M$-theory on $(\widetilde{\mathbb{C}^{2}/\mathbb{Z}_{n}}\times T^{3})$ is isomorphic to the moduli space of flat $SL(n, \mathbb{C})$ connections on $T^{3}$.

\subsubsection{$K$-invariant Ricci flat metrics}

We want to consider all of the  $K$-actions on $M_{GH}^{(n)}$ asymptotic to the action on $\mathbb{C}^{2}/\mathbb{Z}_{n}$.  We now investigate how these act on the coordinates of $M_{GH}^{(n)}$. Quotients of Gibbons-Hawking spaces were also considered in the papers \cite{wright2012quotients, csuvaina2012ale}.
The $G_{2}$-structure on $T^{3}\times M_{GH}^{(n)}$ is given by,
\begin{equation}
\varphi = dy^{1}dy^{2}dy^{3} + d\vec{y}\cdot\vec{\omega},
\end{equation}
and we require that this is $K$-invariant. From the first term we see that $K$ must act on $T^{3}$ such that $d\vec{y}\mapsto M\,d\vec{y}$ where $M\in\op{SL}(3,\mathbb{R})$ and from the second term the action on $M_{GH}^{(n)}$ must induce an action $\vec{\omega}\mapsto N\,\vec{\omega}$ such that $M^{T}=N^{-1}$. If we take the action on $M_{GH}^{(n)}$ to be $(t,\vec{x})\mapsto (t',L\,\vec{x})$ then, using equations \eqref{eq:GH HK forms} and \eqref{eq:potential and connection condition}, we see first that $V(\vec{x})$ must be preserved which means $L\in \op{O}(3)$ and $L$ preserves the set of centres $\{a_{\gamma}\}$ (as discussed below) and second that $A\mapsto \op{det}(L)A$ and $t'=\op{det}(L)\,t$. Then a calculation shows that $\vec{\omega}\mapsto \op{det}(L)\, L\, \vec{\omega}$. Thus $N=\op{det}(L)\, L$ and so $M,N\in\op{SO}(3)$. So even if the action on $\vec{x}$ is in $\op{O}(3)$ the action on the $T^{3}$ and the K\"ahler forms is always in $\op{SO}(3)$. In this paper, however, we will restrict to the cases where the action on the $\mathbb{R}^{3}$ coordinates is in $\op{SO}(3)$, leaving the remaining $\op{O}(3)$ cases for future investigation.

The condition that $V(\vec{x})$ must remain invariant gives an important insight.
Since
\begin{equation}\label{eq:perm}
|L\cdot\vec{x}-\vec{a}_{\gamma}| = |\vec{x}-L^{T}\cdot\vec{a}_{\gamma}|
\end{equation}
whenever $L$ is orthogonal, we see that the action of $K$ on the coordinates $x^i$ is equivalent to an action on the centres. Therefore, in order for $V(\vec{x})$ to be invariant, the elements of $K$ must permute the centres amongst themselves. This is the key restriction on the moduli space that we were seeking. The set of compatible $K$-actions are given by the set of homomorphisms $\chi$ from
\begin{equation}
    \chi : K \mapsto S_{n}
\end{equation}
Some comments are now in order. As explained in \cite{joyce1998topology}, non-identity elements of $S_{n}$ are actually diffeomorphisms of  
$M^{(n)}_{\text{GH}}$ which are \textit{disconnected} from the identity. This is related then to the distinct topologies that can arise on the corresponding $G_2$-manifolds. This is also then clearly related to the distinct components of the moduli space of flat connections on $L$. The insight gained from these examples allows us to address the general case: since the moduli space of ALE metrics on $\widetilde{\mathbb{C}^{2}/\Gamma_{ADE}}$ is given by 
$(\mathfrak{h}_{ADE}\otimes \mathbb{R}^{3})/\op{Weyl}(ADE)$, we must look for homomorphisms from $K$ to the Weyl group. Actually, this is not quite the full answer as, in addition to the action of the Weyl group, the ADE Lie algebras $\mathfrak{g}_{ADE}$ also admit outer automorphisms in general and these could also be induced by actions of $K$, hence the final answer is given by

\begin{equation}
    \chi: K\to \op{Aut}(\Delta_{ADE})\ltimes \op{Weyl}(ADE)
\end{equation}
where $\Delta_{ADE}$ is the Dynkin diagram associated to the $ADE$ Lie algebra.

\subsection{Further Explicit Examples}\label{sec:4.4}

\subsubsection*{Example 1}

Our first example is to consider gauge group $\op{SU}(2)$ and $K=\mathbb{Z}_{2}$ i.e. $X_o(\Gamma_{A_{1}}, \mathbb{Z}_{2})$. 
Let us consider therefore the most general $\mathbb{Z}_{2}$-invariant 2-centre Gibbons-Hawking metric on $\widetilde{\mathbb{C}^{2}/\Gamma_{A_{1}}}$. This amounts to finding all actions of $K=\mathbb{Z}_{2}$ which act as 
\begin{equation}
    K: (\omega_1, \omega_2, \omega_3) \longrightarrow (- \omega_1, - \omega_2, \omega_3)
\end{equation}
on the K\"ahler forms and preserve $V(x_1, x_2, x_3)$. 
In this case we see that the action on the coordinates is
\begin{equation}
    K: (x_1, x_2, x_3) \longrightarrow (-x_1 , -x_2, x_3)
\end{equation}
which, because of \eqref{eq:perm}, is equivalent to the corresponding action on the centres $\vec{a_1}$ and $\vec{a_2}$ which appear in the Gibbons-Hawking potential. We are free to fix the sum of the two centres, $\vec{a_2}=-\vec{a_1}$. We then see that there are two branches to the moduli space of $K$-invariant hyperK\"ahler metrics, given by 
\begin{equation}
    \vec{a_1}=-\vec{a}_{2}=\begin{pmatrix}
        0\\0\\c
    \end{pmatrix}
\end{equation}
for any $c\in\mathbb{R}$
and

\begin{equation}
    \vec{a_1}=-\vec{a}_{2}=\begin{pmatrix}
        a\\b\\0
    \end{pmatrix}
\end{equation}
for any $a,b \in\mathbb{R}$.
In the first case we see that the two centres lie along the fixed point set of the $K$-action on $\mathbb{R}^{3}$
and hence $H_2(\widetilde{\mathbb{C}^{2}/\Gamma_{A_{1}}})$ is preserved by $K$. This corresponds to the Coulomb branch solution. In the second case the two centres are permuted by $K$ and hence the action of $K$ on the homology is non-trivial. This corresponds to the Higgs branch, which we happily see is two-dimensional in this case, in agreement with our gauge theory result from subsection \ref{sec:4.1.1}.

We can also count the $L_{2}$-normalisable harmonic forms on the smooth manifolds $X_{a}(\Gamma_{A_{1}}, \mathbb{Z}_{2})$ (where $a=c$ corresponds to the Coulomb branch and $a=h$ to the Higgs branch) by considering the action of $K$ on the harmonic 2-form $\Gamma_{1}$, defined in \ref{sec:GH space}. In general, since the $K$-action permutes the set of centres, one can see that it also permutes the set of 2-forms $\Omega_{i}$ and thus acts on the $\Gamma_{i}$ as some invertible linear transformation. In this case, on the Coulomb branch $\Gamma_{1}$ is $K$-invariant and on the Higgs branch,
\begin{equation}
    \mathbb{Z}_{2}:\begin{pmatrix}
        \Omega_{1}\\ \Omega_{2}
    \end{pmatrix}\mapsto \begin{pmatrix}
        \Omega_{2}\\ \Omega_{1}
    \end{pmatrix}
\end{equation}
and so $\Gamma_{1}\mapsto -\Gamma_{1}$. Thus we see that $X_{c}$ has $b^{2}=1$ and $b^{3}=1$ and $X_{h}$ has $b^{2}=0$ and $b^{3}=2$ which give rise to the expected number of scalar field moduli and $\op{U}(1)$ factors in the gauge group in the 4d theory arising from compactifying M-theory on $X_{a}(\Gamma_{A_{1}}, \mathbb{Z}_{2})$.

\subsubsection*{Example 2}

This is the example corresponding to $\op{SU}(2)$ gauge theory on $T^{3}/(\mathbb{Z}_{2} \times \mathbb{Z}_{2} )$ i.e. $M$-theory on $X_o(\Gamma_{A_{1}}, \mathbb{Z}_{2} \times \mathbb{Z}_{2})$. The action of $K$ on the $\mathbb{R}^{3}$ coordinates of the Gibbons-Hawking metric is generated by the diagonal order two matrices in $\op{SO}(3)$ of the form:
\begin{equation}
    \beta:=(-1,-1,1)\;\;\;\gamma:=(-1,1,-1)
\end{equation}
In this case there is no Ricci flat metric in which the action of $K$ preserves the homology. Instead there are three components to the moduli space in which the two centres lie along each of the three coordinate axes:
\begin{equation}
    \vec{a_1} = -\vec{a_2} = \begin{pmatrix}
        a\\0\\0
    \end{pmatrix},\quad \begin{pmatrix}
        0\\b\\0
    \end{pmatrix},\quad\begin{pmatrix}
        0\\0\\c
    \end{pmatrix}
\end{equation}
and this is in perfect agreement with the three Higgs branches to the moduli space of $\op{SU}(2)$ flat connections on $T^{3}/K$ from subsection \ref{sec:4.2.2}.

Here, as in the previous example, we can count the harmonic forms. Looking at the first branch of the moduli space, we can compute that both $\beta$ and $\gamma$ act as $\Gamma_{1}\mapsto -\Gamma_{1}$ and so $\beta\gamma$ acts trivially. Thus by wedging $\Gamma_{1}$ with the harmonic 1-form on $T^{3}$ that is odd under $\beta$ and $\gamma$ but even under $\beta\gamma$ we get a $(\mathbb{Z}_{2}\times \mathbb{Z}_{2})$-invariant harmonic 3-form. Thus we have $b^{2}=0$ and $b^{3}=1$ for the desingularisation of $X_o(\Gamma_{A_{1}}, \mathbb{Z}_{2} \times \mathbb{Z}_{2})$ corresponding to this branch and the same for the other two.

\subsubsection*{Example 3}

Let's now consider $\op{SU}(3)$ gauge theory on $T^3/\mathbb{Z}_{2}$, so $\mathbb{Z}_{2}$-invariant 3-centre Gibbons Hawking metrics.
Denote the centres as,
\begin{align}
\begin{pmatrix}
    a_1\\ b_1\\ c_1
\end{pmatrix}, 
\begin{pmatrix}
    a_2\\ b_2\\ c_2
\end{pmatrix},
\begin{pmatrix}
    -a_1-a_2\\ -b_1-b_2\\ -c_1-c_2
\end{pmatrix}
\end{align}
Since $K$ is generated by $g_{\beta}$ and preserves points of the form $(0,0,x_3)$, we have $K$-invariant metrics if the three centres are all of this form. This is the two-dimensional Coulomb branch of the $\mathcal{N}=2$ supersymmetric $\op{SU}(3)$ gauge theory. There are also solutions of the form,
\begin{align}
\begin{pmatrix}
    a\\ b\\ c
\end{pmatrix}, 
\begin{pmatrix}
    -a\\ -b\\ c
\end{pmatrix},
\begin{pmatrix}
    0\\ 0\\ -2c
\end{pmatrix}
\end{align}
This corresponds to an action of $K=\mathbb{Z}_{2}$ which reverses the orientation of the $S^2$ corresponding to the line segment joining the first two centres, but preserves the orientation of the other $S^2$. Hence the Betti numbers of $X_h(\mathbb{Z}_{3}, \mathbb{Z}_{2})$ are $b^{2}=1$ and $b^{3}=3$, the latter corresponding to the three parameters above. 

We can see this three dimensional moduli space in the $\op{SU}(3)$ gauge theory explicitly. Notice that when $a=b=0$ the first two centres coincide and hence an $A_{1}$-singularity appears and that this solution intersects the Coulomb branch there. At this point the physical theory has an unbroken $\op{SO}(2)$ gauge symmetry as the $S^{2}$ which appears when $a$ and $b$ are non-zero is $\mathbb{Z}_{2}$-odd. There is also an unbroken $\op{U}(1)$ gauge symmetry coming from the $S^2$ which connects the first two centres with the third. The corresponding three dimensional moduli space of flat connections is given by,

\begin{equation}\label{SU3 solution}
g_{1}\mapsto\begin{pmatrix}
\lambda_a&0&0\\0&\lambda_a^{-1}&0\\0&0&1
\end{pmatrix},\; g_{2}\mapsto\begin{pmatrix}
\lambda_b&0&0\\0&\lambda_b^{-1}&0\\0&0&1
\end{pmatrix},\; g_{3}\mapsto\begin{pmatrix}
-\lambda_c&0&0\\0& -\lambda_c&0\\0&0&\lambda_c^{-2}
\end{pmatrix},\;  g_{\beta}\mapsto\begin{pmatrix}
0&\lambda_c^{1/2}&0\\ -\lambda_c^{1/2}&0&0\\0&0&\lambda_c^{-1}
\end{pmatrix}
\end{equation}
We see that at the origin of the moduli space the gauge group is $\op{SO}(2)\times \op{U}(1) \subset \op{SU}(2) \times \op{U}(1) \subset \op{SU}(3)$, where the $\op{U}(1)$ factor is in the direction of the Lie algebra which generates $g_3$. Since $g_3$ commutes with $g_1, g_2$ and $g_\beta$, this $\op{U}(1)$ is unbroken for all values of the $\lambda_i$.  We propose that
the massless hypermultiplet which appears at the origin is in the representation $\bf{2_0}$ i.e. a doublet which is neutral under $\op{U}(1)$. Along the Higgs branch the $\op{SO}(2)$ vector multiplet combines with four real bosonic (plus fermionic) degrees of freedom from the hypermultiplet to become a long massive vector multiplet, leaving a massless spectrum consisting of a $\op{U}(1)$ vector multiplet plus a single neutral hypermultiplet. Thus the low energy moduli space is (1+2)-complex dimensional corresponding exactly to the three parameters, $(\lambda_a , \lambda_b, \lambda_c)$ in the family of flat connections. 

For this case we have two harmonic 2-forms, $\Gamma_{1}$ and $\Gamma_{2}$. On the Coulomb branch both 2-forms are invariant and so $b^{2}=b^{3}=2$ for $X_{c}(\mathbb{Z}_{3}, \mathbb{Z}_{2})$ matching the expected two-dimensional moduli space and $U(1)^{2}$ unbroken gauge symmetry at generic points. On the Higgs branch, the action on the 2-forms is,
\begin{equation}
    \begin{pmatrix}
        \Gamma_{1}\\ \Gamma_{2}
    \end{pmatrix}\mapsto \begin{pmatrix}
        -\Gamma_{1}\\ \Gamma_{1}+\Gamma_{2}
    \end{pmatrix}
\end{equation}
and so there is one $\mathbb{Z}_{2}$-odd 2-form, $\Gamma_{1}$, and one $\mathbb{Z}_{2}$-even 2-form, $\Gamma_{1}+2\Gamma_{2}$. Thus, for $X_{h}(\mathbb{Z}_{3}, \mathbb{Z}_{2})$ we have that $b^{2}=1$ and $b^{3} = 3$, as anticipated (explicitly, if the action on $T^{3}$ is  $\alpha: \;(y^{1},y^{2},y^{3})\mapsto (-y^{1},-y^{2},y^{3}+1/2)$, then the harmonic 3-forms are $dy^{1}\wedge \Gamma_{1},\quad dy^{2}\wedge \Gamma_{1},\quad dy^{3}\wedge (\Gamma_{1}+2\Gamma_{2})$).

\subsubsection{The general case for $n$ centres}\label{sec:branches of the mod space}

We will now describe the most general $K$-invariant Gibbons-Hawking Ricci flat metrics by simply describing $K$-invariant configurations of $n$ centres in $\mathbb{R}^{3}$ for arbitrary $n$. 

Firstly, we take $K=\mathbb{Z}_{2} = \langle \alpha\rangle$. Then either one can place $P_{1}$ centres in the $\alpha$-invariant subspace of $\mathbb{R}^{3}$ (such that their centre of mass is at the origin) or one can arrange $P_{2}$ pairs of centres to be exchanged under the action of $\alpha$. So $P_{1} + 2P_{2} = n$ and the dimension of the branch of the moduli space is $P_{1}+3P_{2}-1$. The dimension is this because the centres that are $\alpha$-invariant contribute 1 parameter each (their position on the axis of rotation) and each pair of centres exchanged under $\alpha$ contributes 3 parameters (since such pairs will be of the form $\{(a,b,c), (-a,-b,c)\}$) and the centre of mass condition takes away 1 parameter. Secondly, take $K=\mathbb{Z}_{2}\times \mathbb{Z}_{2} = \langle\alpha, \beta \rangle$. We can place $P_{1}$ centres in the $K$-invariant subspace, in this case this is just the origin so this is the same as considering the case for $n-P_{1}$ centres. We can arrange $P_{2}$ pairs of centres to live in the invariant subspace of, say, $\alpha$ and be exchanged in pairs under $\alpha$ and $\alpha\beta$ (plus cyclic permutation of $\alpha$, $\beta$ and $\alpha\beta$). Or we can arrange centres in $P_{3}$ sets of four as a full orbit of the group. Then $2P_{2}+4P_{3}=n$ and the dimension of the moduli space is $P_{2}+3P_{3}$.

One could continue this reasoning for more general $K$. Thus we can find the number of branches simply by finding all the ways of partitioning the integer $n$ into the tuple $(P_{1},P_{2},...)$ subject to a linear condition on the $P_{i}$ determined by the fact that the centres must be $K$-invariant. As an example, consider $n=4$ and $K=\mathbb{Z}_{2}$. We have 3 branches,
\begin{table}[h]
    \centering
    \begin{tabular}{c|c}
        $(P_{1},P_{2})$ & $d=P_{1}+3P_{2}-1$ \\
        \hline
        $(4,0)$ & $3$
        \\
        $(2,1)$ & $4$
        \\
        $(0,2)$ & $5$
    \end{tabular}
    \label{tab:branches for SU(4)}
\end{table}

\noindent We can explicitly see these branches from the point of view of the centres,
\begin{align}
    &(4,0)\leftrightarrow\left\{\vec{a}_{1} = \begin{pmatrix}0\\0\\c_{1}\end{pmatrix},\;\vec{a}_{2} = \begin{pmatrix}0\\0\\c_{2}\end{pmatrix},\;\vec{a}_{3} = \begin{pmatrix}0\\0\\c_{3}\end{pmatrix},\;\vec{a}_{4} = \begin{pmatrix}0\\0\\-c_{1}-c_{2}-c_{3}\end{pmatrix}\right\}\\
    &(2,1)\leftrightarrow\left\{\vec{a}_{1} = \begin{pmatrix}a_{1}\\b_{1}\\c_{1}\end{pmatrix},\;\vec{a}_{2} = \begin{pmatrix}-a_{1}\\-b_{1}\\c_{1}\end{pmatrix},\;\vec{a}_{3} = \begin{pmatrix}0\\0\\-c_{1}+d_{1}\end{pmatrix},\;\vec{a}_{4} = \begin{pmatrix}0\\0\\-c_{1}-d_{1}\end{pmatrix}\right\}\\
    &(0,2)\leftrightarrow\left\{\vec{a}_{1} = \begin{pmatrix}a_{1}\\b_{1}\\c_{1}\end{pmatrix},\;\vec{a}_{2} = \begin{pmatrix}-a_{1}\\-b_{1}\\c_{1}\end{pmatrix},\;\vec{a}_{3} = \begin{pmatrix}a_{3}\\b_{3}\\-c_{1}\end{pmatrix},\;\vec{a}_{4} = \begin{pmatrix}-a_{3}\\-b_{3}\\-c_{1}\end{pmatrix}\right\}
\end{align}
and we can use the map described in subsection \ref{sec:GH mod space and flat conn on T3} to find the corresponding flat connections, thus giving a prediction for the number and dimension of the branches of the moduli space of flat $\op{SU}(n)$ connections on $T^{3}/K$.

\subsection{Some Higher Rank Examples}

In this section we describe some higher rank $ADE$ Higgs branch solutions by embedding the basic $\op{SU}(2)$ flat connection on $T^{3}/\mathbb{Z}_{2}$ of section \ref{sec:4.1.1} into higher rank groups.

For instance consider the maximal subgroup $\op{SU}(2)^N$ of $\op{SU}(2N)$ (or similarly for $\op{SU}(2N+1)$). We can take $N$ diagonal copies of the $\op{SU}(2)$ solution.
The gauge group at the origin of the Higgs branch for the space,
\begin{equation}
    \frac{T^{3}\times M_{GH}^{(2N)}}{\mathbb{Z}_{2}}
\end{equation}
is the centraliser of $N$ copies of $g_{\beta}$ from \eqref{Higgs solution} as a subgroup of $\op{SU}(2N)$. This subgroup is,
\begin{equation}
\op{S}(\op{U}(N)\times \op{U}(N))\cong (\op{SU}(N)\times \op{SU}(N)\times \op{U}(1))/\mathbb{Z}_{N},
\end{equation}
The fields in the higher dimensional theory (i.e. before compactification onto $T^{3}/\mathbb{Z}_{2}$) transform in the adjoint of $\op{SU}(2N)$. So in order to see how the hypermultiplets in the lower dimensional theory transform we must look at the decomposition of this representation,
\begin{align*}
\op{SU}(2N)&\to (\op{SU}(N)\times \op{SU}(N)\times \op{U}(1))/\mathbb{Z}_{N}\\
\\
\textbf{4N}^{2}-\textbf{1} &\to (\textbf{N}^{2}-\textbf{1},\textbf{1})_{0}+(\textbf{1},\textbf{N}^{2}-\textbf{1})_{0} + (\textbf{N},\bar{\textbf{N}})_{2} + (\bar{\textbf{N}},\textbf{N})_{-2} + (\textbf{1},\textbf{1})_{0}
\end{align*}
which is the adjoint plus the bifundamental and its complex conjugate. The $\mathcal{N}=2$ vector multiplet will transform in the adjoint and the two hypermultiplets in the bifundamentals, thus we have $8N^{2}$ real scalars (the hypermultiplets contain $8$ real scalar degrees of freedom and the dimension of the representation is $N^{2}$). Now we can move away from the origin of the moduli space by switching on VEVs. If we turn on the $(\textbf{N},\bar{\textbf{N}})_{2}$ field we break the $\op{U}(1)$ and the two $\op{SU}(N)$'s break to a diagonal subgroup and $4N^{2}$ of the real scalars become massive. The matter representation breaks to adjoints and singlets of this new group. If we give a VEV one of the remaining scalars in the adjoint we break the group to the maximal torus $\op{U}(1)^{N-1}$, $4(N^{2}-N)$ scalars become massive and we can break no further. So we are left with a $\op{U}(1)^{N-1}$ gauge group and $8N^{2}-4N^{2}-4(N^{2}-N) = 4N$ massless real scalars, these are the moduli on the Higgs branch and match the number of moduli in the explicit flat connection.

This corresponds to $(N-1)$ $\mathbb{Z_2}$-even spheres (giving $N-1$ gauge bosons) and $N$ odd spheres (giving $2N$ \emph{complex} scalars). Indeed, several such $\mathbb{Z}_{2}$ actions can be shown to exist when the centres are arranged in a regular polygon, for example. One can also embed the Coulomb branch solution diagonally with the Higgs branch solution to obtain a mixed branch, which would correspond to having more even spheres. We can achieve this by placing some of the centres on the line that is invariant under the $\mathbb{Z}_{2}$ action.

Also we may explore the moduli spaces of $D$ and $E$ type singularities from the flat connections viewpoint. For the $D$ case this comes from the maximal subgroup of $\op{Spin}(4N)$,
\begin{equation}
    \op{Spin}(4N)\to \op{Spin}(4)^{N}\cong\op{SU}(2)^{2N}
\end{equation}
and then the centraliser of the diagonally embedded $\op{SU}(2)$ Higgs branch solution at the origin is,
\begin{equation}
    \op{Spin}(2N)\times \op{Spin}(2N).
\end{equation}
Away from the origin the gauge group gets completely broken and there are 8$N$ massless real scalars remaining. Similarly, for the $\op{SU}(2)^8$ subgroup of $E_8$ one should get that the surviving gauge symmetry at the origin of the Higgs branch is,
\begin{equation}
    \op{Spin}(16)/\mathbb{Z}_{2} \subset E_8.
\end{equation}
and again, away from the origin the gauge group is completely broken and there are 32 massless real scalars corresponding to 8 copies of the moduli of the basic $\op{SU}(2)$ solution.

Another way in which we can generalise is to consider more general twists than $\mathbb{Z}_{2}$. For example, we could consider the space,
\begin{equation}
    \frac{T^{3}\times M_{EH}}{\mathbb{Z}_{3}}
\end{equation}
where the action on the torus is,
\begin{equation}
    (y^{1},y^{2},y^{3})\mapsto (-y^{2},y^{1}-y^{2},y^{3}+1/3)
\end{equation}
Then on $T^{3}/\mathbb{Z}_{3}$ we have one harmonic 1-form, $dx^{3}$, and two `$\mathbb{Z}_{3}$-twisted' harmonic 1-forms, $e^{-2\pi i/3}\,dx^{1}+dx^{2}$ and $ e^{2\pi i/3}\,dx^{1} + dx^{2}$. However, for this example, the $\mathbb{Z}_{3}$ can only act trivially on the 2-sphere in the Eguchi-Hanson space (since there are no non-trivial homomorphisms from $\mathbb{Z}_{3}$ into $\mathbb{Z}_{2}$) and so we will only get a  Coulomb branch. If we replace $M_{EH}$ with $M_{GH}^{(3)}$ then we do get a Higgs branch and the $\mathbb{Z}_{3}$-twisted 1-forms play the role that the $\mathbb{Z}_{2}$-twisted ones did in our previous discussion. 

\subsection{D-type ALE space}

For $D_{n}$ singularities we can consider the moduli space of flat connections by thinking about configurations of centres as we did for the $A_{n}$ case. Far away from the origin, the ALE space that is the resolution of a $D_{n}$ singularity looks essentially like Gibbons-Hawking space modded out by a $\mathbb{Z}_{2}$ action $(t,\vec{x})\mapsto (-t,-\Vec{x})$ so now instead of thinking about centres in $\mathbb{R}^{3}$ we can think about centres in $\mathbb{R}^{3}/\mathbb{Z}_{2}$ or equivalently, centres in $\mathbb{R}^{3}$ along with their $\mathbb{Z}_{2}$ images. 

We no longer require that the centres sum to zero, since this is trivially satisfied by including their $\mathbb{Z}_{2}$ images. So the only constraints are that the set of centres is invariant under the $K$ action and additionally that $K$ acts on the Dynkin diagram defined by the 2-spheres as an element of $\op{Aut}(\Delta_{D_{n}})\ltimes \op{Weyl}(D_{n}) = \mathbb{Z}_{2}\ltimes (\mathbb{Z}_{2}^{n-1}\ltimes S_{n})$ (except for $n=4$, when $\op{Aut}(\Delta_{D_{n}}) = S_{3}$). The $\mathbb{Z}_{2}^{n-1}$ factor is important; it corresponds to multiplying an even number of the coordinates of a $D_{n}$ root vector by $-1$. Geometrically, this corresponds to sending an even number of centres to their $\mathbb{Z}_{2}$ images. Thus, on top of the constraint that $K$ preserves the set of centres, we also have that if $K$ sends $m$ centres to their $\mathbb{Z}_{2}$ images, $m$ must be even (note that for the $A$-type examples the fact that $K$ preserves the set of centres automatically means it acts as an element of the Weyl group on the 2-spheres, so we got no further constraint like we do here). 

As an example, consider $D_{2}$ with $K=\mathbb{Z}_{2}$ \footnote{There is no ALE space for this example but there is an ALF space. For our arguments this difference is irrelevant.}. We expect this to coincide with two  copies of our $A_{1}$ example by virtue of the isomorphism $\mathfrak{so}(4)\cong \mathfrak{su}(2)\times \mathfrak{su}(2)$. The allowed configurations of centres (omitting the orientifold images) are,
\begin{align}
    &\left\{\vec{a}_{1} = \begin{pmatrix}0\\0\\c_{1}\end{pmatrix},\;\vec{a}_{2} = \begin{pmatrix}0\\0\\c_{2}\end{pmatrix}\right\}\\
    &\left\{\vec{a}_{1} = \begin{pmatrix}a_{1}\\b_{1}\\c_{1}\end{pmatrix},\;\vec{a}_{2} = \begin{pmatrix}-a_{1}\\-b_{1}\\c_{1}\end{pmatrix}\right\}\\
    &\left\{\vec{a}_{1} = \begin{pmatrix}a_{1}\\b_{1}\\0\end{pmatrix},\;\vec{a}_{2} = \begin{pmatrix}a_{2}\\b_{2}\\0\end{pmatrix}\right\}
\end{align}
Where the first branch corresponds to two copies of the $A_{1}$ Coulomb branch, the second to a one copy of the $A_{1}$ Coulomb branch and one copy of the Higgs branch and the third to two copies of the $A_{1}$ Higgs branch. By looking at the corresponding flat connection, we see that on the first branch the full $\op{SO}(4)$ is unbroken at the origin of the branch and at generic points this is broken to $\op{SO}(2)^{2}$, on the second branch there is an unbroken $\op{SO}(2)\times \op{SU}(2)$ at the origin and at generic points an unbroken $\op{SO}(2)$ and on the third branch there is an unbroken $\op{SO}(2)\times \op{SO}(2)$ at the origin which is completely broken at generic points. This matches nicely with what we would expect from our considerations for $A_{1}$.

Next, we consider $D_{3}$ with $K=\mathbb{Z}_{2}$. This example  should give the same answer as the $A_{3}$ example, thanks to the  isomorphism $\mathfrak{so}(6)\cong \mathfrak{su}(4)$. The allowed configurations of centres are,
\begin{align}
    &\left\{\vec{a}_{1} = \begin{pmatrix}0\\0\\c_{1}\end{pmatrix},\;\vec{a}_{2} = \begin{pmatrix}0\\0\\c_{2}\end{pmatrix},\;\vec{a}_{3} = \begin{pmatrix}0\\0\\c_{3}\end{pmatrix}\right\}\\
    &\left\{\vec{a}_{1} = \begin{pmatrix}a_{1}\\b_{1}\\c_{1}\end{pmatrix},\;\vec{a}_{2} = \begin{pmatrix}-a_{1}\\-b_{1}\\c_{1}\end{pmatrix},\;\vec{a}_{3} = \begin{pmatrix}0\\0\\c_{2}\end{pmatrix}\right\}\\
    &\left\{\vec{a}_{1} = \begin{pmatrix}a_{1}\\b_{1}\\0\end{pmatrix},\;\vec{a}_{2} = \begin{pmatrix}a_{2}\\b_{2}\\0\end{pmatrix},\;\vec{a}_{3} = \begin{pmatrix}0\\0\\c_{3}\end{pmatrix}\right\}
\end{align}
where, using the notation of subsection \ref{sec:4.4}, the first branch corresponds to the $A_{3}$ branch labelled $(4,0)$, the second branch to $(2,1)$ and the third to $(0,2)$ and dimensions of the branches agree as expected. Note how, for example, the 6d solution given by $\left\{\vec{a}_{1} = (a_{1},b_{1},0),\;\vec{a}_{2} = (a_{2},b_{2},0),\;\vec{a}_{3} = (a_{3},b_{3},0)\right\}$ is not allowed since the $K=\mathbb{Z}_{2}$ action sends an odd number of centres to their images. 

\subsection{Compact Examples}

Finally, we briefly also discuss compact examples, where now the low energy theory is a four dimensional supergravity theory coupled to vector multiplets and chiral multiplets. The general strategy is clear: the chiral multiplet representations under the $ADE$-gauge symmetries are determined by the Weyl group action and flat connections as above.

For the basic Joyce-Karigiannis examples, the answer (as given in section 3) is clear when one considers the $\mathbb{Z}_{2}$-twisted desingularisation of an $A_{1}$-singularity: one obtains an $\op{SO}(2)$ gauge theory with a chiral multiplet in the fundamental representation. We point out that several of Joyce's original examples are of this form \cite{joyce1996compact2}.

\subsubsection{Joyce Examples}

All of the examples in \cite{joyce1996compact2, dominic} are obtained by considering a quotient of a flat 7-torus by a finite group, $\Gamma$ producing an orbifold whose singularities, in favourable cases can be resolved, producing a smooth 7-manifold with metrics of $G_2$-holonomy.
In \cite{joyce1996compact2}, in all except two examples (17 and 18), all of the singularities that occur are of the kinds described in this paper since they are all locally modelled on $X_0(\Gamma_{ADE}, K)$. Hence, the results presented here allow one to simply read off the key ingredients of the low energy field theory arising from those examples.
For instance, all of the examples of Table 1 of \cite{joyce1996compact2} have $A_{1}$-singularities fibered over $T^3$ or $T^3/{\mathbb{Z}_{2}}$. Hence for each $T^3$ one has an $\op{SU}(2)$ vector multiplet plus three adjoint chiral multiplets. Each $T^3/\mathbb{Z}_{2}$ instead gives rise either to an $\op{SO}(2)$ vector multiplet and a doublet of chiral multiplets or to an $\op{SU}(2)$ vector multiplet and a single adjoint chiral multiplet depending on which choice of resolution one makes.

Higher order singularities of the form $X_0(\mathbb{Z}_{3}, \mathbb{Z}_{2})$
also occur, e.g. in example 15 of \cite{joyce1996compact2} the singular set contains one component with singularity modelled on $X_0(\mathbb{Z}_{3}, \mathbb{Z_2})$. As discussed in section \ref{sec:4.4}, there are two distinct resolutions of $X_0$, one which gives rise to an $\op{SU}(3)$ vector multiplet plus an adjoint chiral multiplet (and a two-dimensional space of Coulomb vacua) and a second which gives rise to an $\op{SO}(2)\times \op{U}(1)$ vector multiplet with a pair of chiral multiplets both in the fundamental of $\op{SO}(2)$ (and a three-dimensional space of vacua).

Joyce's examples were extended by Barrett \cite{Barrett} and Reidegeld \cite{Reidegeld} and include cases containing $A_n$, $D_n$ and $E_6$ singularities fibered over $T^{3}/K$ for various $K$. The corresponding gauge and matter representations can similarly be obtained from the results presented here \cite{Baldwin}.

\bigskip
\large
\noindent
{\bf {\sf Acknowledgements.}}
\normalsize

We would like to thank R. Barbosa, L. Foscolo, D. Joyce, S. Karigiannis and J. Lotay for discussions.
The work of BSA and DB is supported by a grant from the Simons Foundation (\#488569, Bobby Acharya)

\bigskip

\bibliographystyle{plain} 

\bibliography{References}

\end{document}